# A Multi-model and Multi-scenario Assessment of the Impact of Climate Change on the Heating and Cooling Load Components of an Archetypical Residential Room in Major Indian Cities

Raj S. Srivastava [a, b #], A. Aravinda De Chinnu [a, #], and Aakash C. Rai [a, b, c *]

[a] Department of Mechanical Engineering, Birla Institute of Technology and Science, Pilani, Rajasthan, India - 333031.

[b] Chandrakanta Kesavan Centre for Energy Policy and Climate Solutions, Indian Institute of Technology Kanpur, Uttar Pradesh, India - 208016.

[c] Department of Sustainable Energy Engineering, Indian Institute of Technology Kanpur, Uttar Pradesh, India - 208016.

[#] Both authors contributed equally to this work.

[*] Corresponding author: Email: aakashrai@iitk.ac.in

**Abstract**

Residential heating and cooling (H/C) currently account for ~7% of India's electricity consumption. A warming climate will increase residential cooling requirements, while heating needs will decrease — an alarming consequence for India, which has predominantly cooling requirements. Thus, to reduce the Indian building sector's energy and carbon footprint, it is essential to assess the impact of climate change on future H/C needs and develop energy-efficiency solutions. This research evaluated the effect of climate change on the H/C energy needs of an archetypical residential room in India, covering all climate zones. We developed a novel approach to quantify the H/C load components (walls, windows, etc.) for identifying building elements to be targeted for improving energy efficiency. Our median climate model predicted that the cooling energy demand of the room archetype would increase by 23–155% by the 2090s compared to the 1990s, depending on the city and emission scenario. Walls and windows account for over 60% of the cooling needs and should be the prime targets for energy-efficiency measures. We also predict between 45–100% reductions in the heating requirements of the archetype room by the 2090s. Walls contribute over 67% to heating needs and should be targeted for heating energy reductions.

**Nomenclature**

| | | | |
|---|---|---|---|
| $A$ | area, m$^2$ | *Greek symbol* | |
| | | $\rho$ | density, kg.m$^{-3}$ |
| $ACH$ | air changes per hour | | |
| $c_p$ | specific heat capacity, J.kg$^{-1}$.°C$^{-1}$ | *Subscripts* | |
| $E_H$ | heating load, kWh | $inf$ | infiltration |
| $h$ | convective heat transfer coefficient, W.m$^{-2}$.°C$^{-1}$ | $int$ | internal load components |
| $h_{fg}$ | latent heat of vaporization of water, J.kg$^{-1}$ | $s$ | internal surfaces |
| $\dot{m}$ | mass flow rate, kg.s$^{-1}$ | $sys, lat$ | system, latent |
| $N$ | number of internal loads | $sys, sen$ | system, sensible |
| $N_{surfaces}$ | number of internal surfaces with convective heat transfer | $Wall$ | exterior walls |
| $\dot{Q}$ | heating/cooling energy, J | $Wall\ Norm$ | exterior walls normalized |
| $T$ | temperature, °C | $Walls$ | actual exterior walls |
| $t$ | time, s | $w$ | moisture |
| $V$ | volume, m$^3$ | $Win$ | windows |
| $W$ | air humidity ratio | $z$ | zone air |

## 1. Introduction

Global warming is one of the most severe problems of the 21$^{st}$ century, due to which the combined land and ocean surface temperature has increased at an average rate of 0.08 °C/decade since 1880 [1]. However, since 1981, the average rate of temperature increase has been more than twice that rate (0.18 °C/decade), which highlights the growing severity of the problem [1]. In fact, the last seven years (2015–21) have been the warmest seven on record [2]. Rising temperatures lead to changes in weather conditions [3] that can significantly impact energy consumption patterns [4]. For example, a warming climate reduces space heating consumption in buildings while increasing the cooling needs. This aspect is particularly troublesome for India — a primarily tropical country with more than three times the cooling needs per person and four times the population of the United States [5]. A warming climate coupled with an increase in the built-up area and income levels will drastically increase India's residential cooling needs, which already account for ~7 % of the country's electricity consumption [6,7] and ~3 % of associated greenhouse gas (GHG) emissions [8,9].

## 1.1. Impact of climate change on the future heating and cooling demand of residential buildings

Several investigations have assessed the role of climate change on the future heating and cooling (H/C) energy demand of residential buildings due to its significant contribution to global energy consumption and GHG emissions. Some studies have used heating and cooling degree-days, which are usually the aggregates of the differences between daily mean temperatures and a base temperature, to predict the current and future H/C energy requirements [10–14]. For example, Isaac and Vuuren [15] used degree days to show that due to climate change, the worldwide residential heating energy demand will decrease by 34 % by 2100, while the cooling energy demand will increase by 72 % over its current value. Similarly, in Indian cities, climate change will decrease the heating energy demand by 19.3–97.1 %, while cooling demand will increase by 11.89–83.0 % in the 2080s [16].

Although degree-days are easy-to-use metrics calculated from outdoor air temperature patterns, they generally do not account for other weather parameters (humidity, solar radiation, etc.) and building characteristics, which can significantly impact the H/C energy consumption [17,18]. Thus, to include building and climate characteristics, some studies have conducted thermal simulations of archetype buildings to get more precise estimates of the H/C energy requirements under current and future climatic conditions (see Table 1). Table 1 shows that several studies have investigated the change in H/C demand under the A2 emission scenario. This scenario is based on the special report on emission scenarios (SRES), and envisions a world with a rising population and relatively slow economic and technological advancements and lies on the higher end of the GHG emission spectrum [19]. Under the A2 scenario, the heating demand in the future is expected to decrease by 14.7–98 %, while the cooling demand will increase by 17–30946 % due to climate change, depending on the city, building type, and time period. Note that there is a huge percentage increase in cooling demand in some studies given in Table 1, due to the low baseline cooling needs.

Since the emission scenarios given in the SRES were superseded by the representative concentration pathways (RCPs) in the fifth assessment report of the Intergovernmental Panel on Climate Change (IPCC), recent studies have investigated the changes in H/C demand under different RCPs. These RCPs are characterized by increased radiative forcing (RF) in 2100 with respect to 1750, primarily due to rising GHG in the atmosphere. As shown in Table 1, the most commonly studied pathway was RCP8.5 (RF = 8.5 W/m$^2$), which is another high GHG emission scenario [20]. Under RCP8.5, the heating demand will decrease by 14–74 %, while the cooling demand will increase by 30–327 %, depending on the location, building type, and time period (see Table 1).

*Table 1: Summary of recent studies assessing the impact of climate change on residential buildings' heating and cooling energy demands (HD and CD).*

| S. No. | Location | Residential building type | Future periods (Baseline year) | Scenario | Decrease in HD (%) | Increase in CD (%) |
|---|---|---|---|---|---|---|
| **Studies based on emission scenarios from SRES** | | | | | | |
| 1. [21] | 3 cities, Brazil | 1-storey house | 2020s, 2050s & 2080s (1961–90) | A2 | 63–98 | 43–400 |
| 2. [22] | 4 cities, USA | 3-storey house | 2050s (1961–90) | A1F1 | 27.4–48.9 | 24.2–36.4 |
| | | | | A2 | 14.7–35.4 | 17.4–27 |
| 3. [23] | 5 cities, Taiwan | 1-storey apartment | 2020s & 2080s (1979–2003) | A1B | – | 24–184 |
| 4. [24] | 2 cities, Brazil | 1-storey house | 2050s (1961–90) | A2 | – | 39–140 |
| 5. [25] | 4 cities, Argentina | 1-storey house | 2080s (1961–90) | A2 | 23–59 | 257–693 |
| 6. [26] | Accra, Ghana | 1-storey house | 2030s & 2050s (2000–2009) | A1B | – | 31–50 |
| 7. [27] | 19 cities, Europe | 3-storey building (9 apartments) | 2050s & 2080s (1961–90) | A2 | 18–82 | 71–30946 |
| 8. [28] | Darwin, Australia | 1-storey house | 2060 (2016) | A2 | – | 52 |
| 9. [29] | Toronto, Canada | 4-storey building (32 apartments) | 2050s (1998–2014) | A2 | 29.5 | 35.5 |
| | | 10-storey building (80 apartments) | | | 28 | 43 |
| 10. [30] | Rome, Italy | 2-storey house | 2050s (1982–99) | A2 | 32 | 169.5 |
| | | 4-storey building | | | 33.9 | 100 |
| 11. [31] | 5 cities, India | 2-storey house | 2020s, 2050s & 2080s (1961–90) | A2 | 21.5–95 | 18.5–185 |
| 12. [32] | Lecce, Italy | 3-storey building (6 apartments) | 2020s, 2050s & 2080s (2010) | A2 | 12.5–61 | 27–119 |
| **Studies based on RCPs from IPCC's fifth assessment report** | | | | | | |
| 1. [33] | Vaxjo, Sweden | 6-storey building (24 apartments) | 2050s & 2090s (1961–90) | RCP4.5 | 22–23 | 76–156 |
| 2. [34] | 74 cities, Chile | 2-storey house | 2060 (2006) | RCP2.6 | 7–16 | – |
| | | | | RCP8.5 | 14–34 | – |
| 3. [35] | Valencia, Spain | 2-storey house | 2050s & 2090s (1961–90) | RCP4.5 | 33.2–41 | 50–103.3 |
| | | | | RCP8.5 | 43.2–66.8 | 100–327 |
| 4. [36] | Hong Kong | 40-storey building (320 apartments) | 2090 (1979–2003) | RCP8.5 | – | 134 |
| 5. [37] | Florence, Italy | 3-storey house | 2050s & 2080s (2000–2015) | RCP8.5 | 30–56 | 33–85 |
| | | 5-storey building (6 apartments) | | | 39–74 | 30–79 |
| 6. [30] | Rome, Italy | 2-storey house | 2050s (1982–99) | RCP8.5 | 30.4 | 162.3 |
| | | 4-storey building | | | 33.8 | 104.1 |
| 7. [29] | Toronto, Canada | 4-storey building (32 apartments) | 2050s (1998–2014) | RCP8.5 | 36 | 41 |
| | | 10-storey building (80 apartments) | | | 36.5 | 51 |
| 8. [38] | 2 cities, USA | 2-storey house | 2050 & 2080 (2020) | RCP2.6 | – | 0–25 |
| | | | | RCP4.5 | – | 18.2–25 |
| | | | | RCP7.0 | – | 18.2–75 |
| | | | | RCP8.5 | – | 27.3–100 |
| | | 10-storey building | | RCP2.6 | – | 3.8–11.7 |
| | | | | RCP4.5 | – | 11.7–18.4 |
| | | | | RCP7.0 | – | 13.4–42.9 |
| | | | | RCP8.5 | – | 21.8–69.1 |

**1.2. Research gap and objectives**

As evident from the above literature review, several studies have assessed the impact of climate change on the H/C energy consumption of residential buildings. Those studies consistently point toward decreasing heating and increasing cooling needs, albeit with significant variations in their estimates due to the differences in building types and climatic conditions. Nevertheless, the following research gaps remain that motivate the present study:

- No study has reported the future projections of H/C load components, which quantify the contribution of different building elements (walls, windows, infiltration, etc.) to the total H/C energy consumption. We developed a novel approach to quantify those load components and estimated them for an archetypical residential room under current and future climatic conditions in eight Indian cities. The method is useful for identifying those elements that should be the focus of energy-efficiency measures and can aid future studies on the topic.

- Previous studies used a single climate model or a single emission scenario to quantify the impact of climate change on building H/C energy demand in India [16,31]. However, future climate and H/C energy forecasts are highly sensitive to the models and scenarios used for making those predictions [39,40]. Thus, the present study conducted a multi-model and multi-scenario assessment of the impact of climate change on the H/C energy consumption in Indian residences to incorporate the associated uncertainties into future projections.

**2. Methodology**

This study performed energy simulations of an archetypical residential room in India under present and future climatic conditions to evaluate the impact of climate change on its H/C load components. To study how residences in different climatic zones would respond to climate change, we selected eight major cities of India, covering all five climate zones of the country, as shown in Figure 1 and Table 2. A description of the selected cities with their climatic conditions is given in Table 2. The simulation methodology and construction of weather files are described in the following sub-sections.

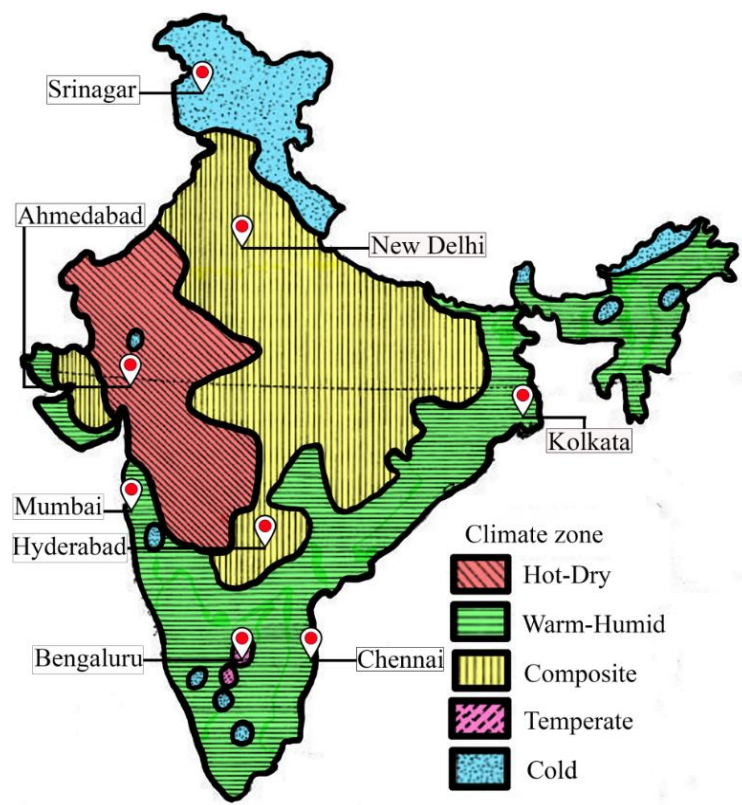

*Figure 1: The selected cities covering all climate zones of India*

*Table 2: Description of the selected cities.*

| City | Description | Climate zone* | Climate zone** |
|---|---|---|---|
| Ahmedabad | A large metropolis located in western India. Winters are short and mild, with the lowest monthly temperature being 12 °C, whereas summers are long and harsh, with maximum monthly temperatures reaching up to 42 °C. | Hot–Dry | BSh |
| Chennai | Coastal cities and major metropolitan hubs of the country. They cities experience a very mild winter season with temperatures generally exceeding 15 °C throughout the day. Summers are hot and humid with maximum monthly temperatures ranging between 34–38 °C. | Warm-Humid | Aw |
| Kolkata | | | Aw |
| Mumbai | | | Am |
| Delhi | India's capital and a major metropolitan area. Winters are cold, with temperatures dropping routinely below 10 °C at night, while summers are hot with temperatures exceeding 40 °C during the day. | Composite | BSh |
| Hyderabad | A major metropolitan city located in southern India. Winters are mild, with temperatures generally ranging between 15–30 °C, while summers are hot with temperatures exceeding 40 °C. | | Aw |
| Bengaluru | A large metropolitan city in southern India that experiences a moderate climate with temperatures generally ranging between 16–34 °C throughout the year. | Temperate | Aw |
| Srinagar | India's most populous city in the cold climate zone that lies in a Himalayan valley. Winters are cold, with temperatures dropping below 0 °C at night, while summers are mild, with the maximum daytime temperature being around 30 °C. | Cold | Cfa |

*According to the climate classification of the Indian Bureau of Energy Efficiency [41], ** According to the Köppen climate classification [42].

**2.1. Building energy simulation**

This investigation conducted energy simulations of an archetypical bedroom, as shown in Figure 2, due to the general Indian practice of using heating and cooling systems in bedrooms only. The archetypical bedroom represents an affluent urban household that uses a room air-conditioner for cooling and an electric resistance heater for heating. Note that only 8% of Indian households have access to room air-conditioning; however, this is projected to grow to 40% by the mid-2030s [43]. Thus, the archetype room represents wealthy urban households, which have the largest share of the country's household electricity consumption and GHG emissions [44].

The room archetype was adapted from the design guidelines provided by the Indian Bureau of Energy Efficiency [45,46]. The room (3.33 m × 4.03 m × 3.18 m in size) was assumed to be on an intermediate floor with two exterior walls. The room's exterior walls had glass windows with static drapes and horizontal overhangs at lintel level. The exterior walls were north and east facing for all locations with warm climates, except for Srinagar (cold climate), where they were south and west facing. Those wall orientations were chosen to reduce the cooling energy in warm climatic zones and heating energy in the cold climate zone since north and east facing walls receive less solar radiation than south and west facing walls for the selected locations [45,46]. The details of the envelope material and its thermophysical properties are given in Table SM1.

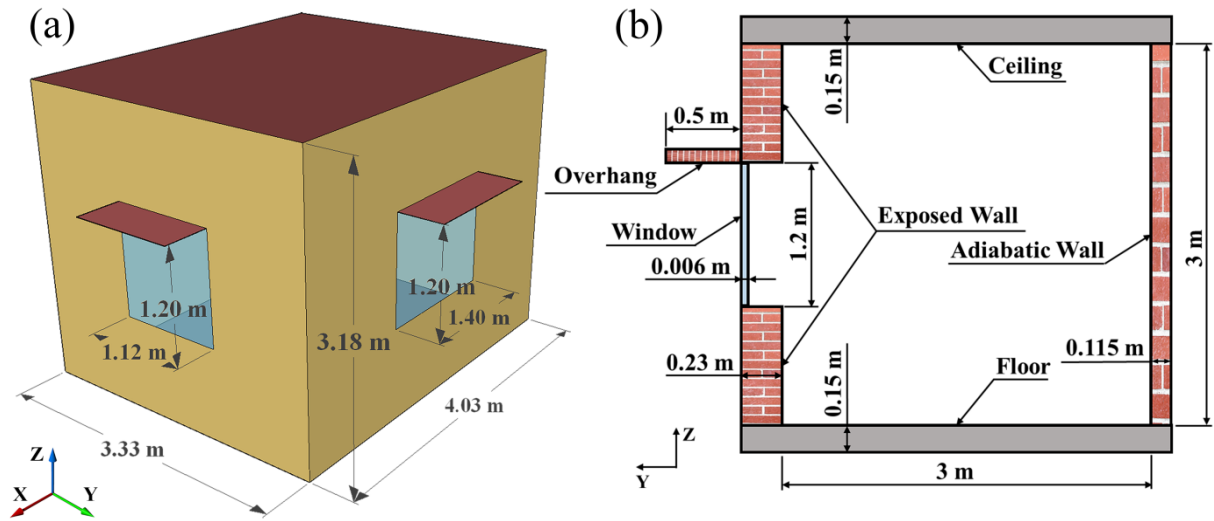

*Figure 2: Schematic of the archetype residential room a) 3-dimensional rendering and b) sectional view.*

### 2.1.1. Boundary conditions and simulation parameters

We performed hourly energy simulations using EnergyPlus software, which has been extensively used and validated for studying buildings' energy performance [47,48]. In our simulations, the room's exterior walls were exposed to outdoor conditions obtained from the current and future weather files, as discussed in Section 2.2. The interior walls, floor, and ceiling were assumed to be adiabatic as the heat transfer through these surfaces would be pretty small since the room was on an intermediate floor [49]. We assumed a constant infiltration rate of 0.75 h$^{-1}$ throughout the year [45,46]. Two occupants were considered present in the room every day from 21:00–07:00 hours, and lights (54 W power) were operational from 21:00–23:00 hours. The H/C system was active only during occupied hours, with a cooling set-point of 26 °C and a heating set-point of 18 °C. Dehumidification was provided when the relative humidity exceeded 65 %; however, humidification was not used since it is not generally employed in Indian households.

### 2.1.2. Heating and cooling load

The H/C system in the room was modeled using the ideal-loads air system in EnergyPlus, which can be considered a perfect variable air volume terminal that removes heat and moisture at 100 % efficiency [47]. We calculated the room's hourly H/C requirements (sensible and latent) and aggregated them to compute the annual H/C needs.

The hourly *sensible* H/C requirement was obtained by using zone air heat balance, which is given by:

$$\dot{Q}_{sys,sen} = \sum_{i=1}^{N} \dot{Q}_i + \sum_{i=1}^{N_{surfaces}} h_i A_i (T_{si} - T_z) + \dot{m}_{inf} c_p (T_{inf} - T_z) - \rho_{air} c_p V_z \frac{dT_z}{dt} \qquad \text{Eq. (1)}$$

where, $\dot{Q}_{sys,sen}$ is the room's sensible heating or cooling requirement, $\sum_{i=1}^{N} \dot{Q}_i$ the sum of internal convective loads (occupants and lights), $\sum_{i=1}^{N_{surfaces}} h_i A_i (T_{si} - T_z)$ the sum of the convective loads from the zone's internal surfaces (walls, ceiling, and floor), $\dot{m}_{inf} c_p (T_{inf} - T_z)$ the sensible part of infiltration load, and $\rho_{air} V_z c_p \frac{dT_z}{dt}$ the rate of change of sensible heat stored in the air.

We calculated the *latent* cooling requirements from the zone air moisture balance, given by:

$$\dot{Q}_{sys,lat} = \sum_{i=1}^{N} \dot{m}_{w,i} h_{fg} + \dot{m}_{inf} h_{fg} (W_{inf} - W_z) - \rho_{air} V_z h_{fg} \frac{dW_z}{dt} \qquad \text{Eq. (2)}$$

where, $\dot{Q}_{sys,lat}$ is the latent cooling energy requirement of the room, $\sum_{i=1}^{N} \dot{m}_{w,i} h_{fg}$ is the sum of latent internal loads (occupants), $\dot{m}_{inf} h_{fg} (W_{inf} - W_z)$ the latent part of infiltration load, and $\rho_{air} V_z h_{fg} \frac{dW_z}{dt}$ the rate of change of latent heat stored in the air.

**2.1.3. Heating and cooling load components**

This study developed a novel procedure to quantify the contributions of building elements such as walls, windows, infiltration, lighting, and occupants to the annual H/C energy consumption, referred to as the "load components". This approach calculates the incremental increase or decrease in the building's H/C energy consumption by adding its components (wall, windows, etc.) one at a time. The increment in the H/C energy is assigned as the load for the corresponding component. Our approach ensures that the sum of all the load components will always be equal to the total load.

We describe the procedure for obtaining the heating load components only since the same steps can be followed for quantifying the cooling load components. To estimate the heating load components, we initially modeled the archetype room with walls only (Model #1), i.e., without windows, infiltration, lighting, or occupants, and calculated the corresponding heating requirements ($E_{H\ Model\ \#1}$ in kWh) using EnergyPlus. Those heating requirements were normalized with the exterior wall area (area = 23.4 m² since windows were not part of Model #1) to determine the walls' contribution to the load per unit area ($E_{H\ Wall\ Norm}$ in kWh/m²) as:

$$E_{H\ Wall\ Norm} = \frac{E_{H\ Model\ \#1}}{A_{Wall\ Model\ \#1}} \qquad \text{Eq. (3a)}$$

Next, we added windows to the model (Model #2) and calculated the heating load ($E_{H\ Model\ \#2}$). The contributions of the walls and windows to the heating loads were calculated from Eqs. 3 b–c, respectively as:

$$E_{H\ Walls} = E_{H\ Wall\ Norm} \times A_{Walls} \qquad \text{Eq. (3b)}$$

$$E_{H\ Win} = E_{H\ Model\ \#2} - E_{H\ Walls} \qquad \text{Eq. (3c)}$$

where $E_{H\ Walls}$ and $E_{H\ Win}$ are the heating load components corresponding to walls and windows, respectively, and $A_{Walls}$ is the actual wall area (20.4 m$^2$).

Subsequently, infiltration and internal heat gains (lighting and occupants) were added to the model, and we calculated the heating requirements in those models ($E_{H\ Model\ \#3}$ and $E_{H\ Model\ \#4}$). Finally, the heating loads attributed to infiltration ($E_{H\ Inf}$) and internal gains ($E_{H\ Int}$) were estimated from Eqs. 3 d–e, respectively:

$$E_{H\ Inf} = E_{H\ Model\ \#3} - E_{H\ Model\ \#2} \qquad \text{Eq. (3d)}$$

$$E_{H\ Int} = E_{H\ Model\ \#4} - E_{H\ Model\ \#3} \qquad \text{Eq. (3e)}$$

Note that a negative load component means that the particular element does not contribute to the load; instead, it reduces it. For example, the internal gains component ($E_{H\ Int}$) of the heating load will be negative.

The load components calculated from our approach depend on the order in which the elements (windows, infiltration, and internal) are added to the building model. To address this issue, we evaluated the current load components by following all six possible orders in which we could add the remaining building elements (windows, infiltration, and internal) to the model containing walls only. Our results showed that the cooling load components had a slight variation (less than 12 percentage points) between the different orders that can be used, as shown in Figures SM1 a–h for the different cities. Similarly, the heating load components estimated for Srinagar were about the same, irrespective of the ordering used (see Figure SM2 a). However, in New Delhi, the heating load components had much larger variations (up to 36 percentage points) as shown in Figure SM2 b, depending on the ordering used due to the relatively small heating requirements there (63 kWh). Thus, the method seems reasonably robust when the loads are significant, and any ordering scheme would yield reasonable estimates.

**2.2. Weather data**

To assess the impact of climate change on the H/C load components, we first estimated them under current weather conditions in the eight cities and then compared them with their future values under two different climate change scenarios. The details of the current and future weather data used in this study are provided in the next subsections. Those future weather files are available to download from: https://github.com/aakashchandrai/Weather-Files.

**2.2.1. Present weather conditions**

We obtained the present weather data for the selected cities from the Typical Year (TY) weather files published by the Indian Society of Heating, Refrigerating, and Air-conditioning Engineers (ISHRAE) [50,51]. ISHRAE's TY files contain hourly weather data for one year and represent typical meteorological conditions at a given location. Those TY files used weather observations during 1984–2008; thus, 1984–2008 (referred to as the 1990s) is the baseline period for this study. We modified the ISHRAE TY files since they used the Zhang model [52] to calculate the direct-normal and diffuse components of the solar radiation. However, we used the Boland–Ridley–Laurent (BRL) model [53] for estimating those components to maintain consistency with the future weather files.

**2.2.2. Future weather projections**

We generated future TY files by first estimating the monthly shifts in weather parameters for future periods (FPs), which refer to the differences between the monthly averaged value of a weather parameter in any FP and the baseline (1990s). Those shifts were estimated for three FPs: 2026–2045 (2030s), 2056–2075 (2060s), and 2081–2100 (2090s) by using outputs from General Circulation Models (GCMs). After obtaining those shifts, we used the morphing technique developed by Belcher et al. [54] to generate future TY files at hourly resolution. The detailed procedure for generating the future TY files is given in the following paragraphs.

We used six different GCMs (see Table SM2) for estimating the monthly weather shifts, that were selected based on the availability of weather parameters required for conducting energy simulations. The GCM weather data was obtained for all the FPs under RCP4.5 (RF = 4.5 W/m$^2$) and RCP8.5 (RF = 8.5 W/m$^2$), which represent medium and high emission pathways, respectively [55]. The weather data was estimated at the selected locations using the inverse-distance-weighting interpolation scheme [56].

To deal with uncertainties associated with weather predictions, the monthly weather shifts were averaged across all available ensemble runs (ensembles are generated using identical radiative forcing but slightly different initial conditions) of the selected GCMs [57]. This procedure generated six different values of the monthly weather shifts (one corresponding to each GCM) corresponding to each city, FP, and RCP. For example, corresponding to the variable temperature, we obtained the following shifts for January month in the 2030s in New Delhi under RCP4.5: 0.71 °C, 0.86 °C, 1.37 °C, 1.49 °C, 1.62 °C, and 1.63 °C. Thus, the minimum projected shift is 0.71 °C, the median is 1.43 °C, and the maximum is 1.63 °C, corresponding to the minimum, median and maximum predicted increase in January month's temperature by the different GCMs. Those minimum, median and maximum values of the temperature shifts were used to generate corresponding model classes: min, median, and max, to account for inter-model variabilities, without generating future files corresponding to all the six GCMs. Further details regarding the min, median, and max models are provided in Section 1 of Supplementary Material (SM). We used the min, median, and max model shifts to generate future TY files for a given location, period, and scenario, as explained in the next paragraph.

From the monthly weather shifts, we constructed the future TY weather files that contain hourly values of the future weather parameters (temperature, humidity, solar radiation, etc.) using the morphing method [54], as given in SM Section 2. This method generates future weather forecasts containing the average conditions of future climate while preserving realistic weather patterns. Since the future TY files also require hourly values of direct-normal and diffuse irradiations, they were estimated from the hourly value of the global horizontal irradiance (GHI) using the BRL model [53]. Finally, we evaluated the dew point temperature and horizontal infrared radiation intensity from the sky using EnergyPlus's weather converter [58].

## 3. Results and Discussion

This section first discusses the changes in climate variables (temperature, humidity, solar radiation, and wind) in the eight Indian cities by comparing the baseline climate with future climate projections. Next, we discuss the current and future H/C energy demand of the archetype residential room in each city. Finally, a detailed analysis of the individual H/C load components is presented to identify the target building elements for improving residential energy efficiency.

### 3.1. Future climate projections

Figure 3 shows the projected increase in the annual-averaged air temperatures in the different cities over the baseline value (in the 1990s) under RCP4.5 and RCP8.5. For all cities, the air temperatures will increase consistently from the 1990s to the 2090s under both emission scenarios, irrespective of the model used (min, median, or max) for making future projections. By the 2030s, the temperature rise would be almost the same for both RCPs, i.e., between 0.8–1.7 °C (median model), depending on the city. By the 2060s, the median model predicts temperature rise to be slightly higher under RCP8.5 (2.0–3.7 °C) than under RCP4.5 (1.4–2.3 °C). However, by the 2090s, the temperature rise would be almost double under RCP8.5 (2.8–5.3 °C) than under RCP4.5 (1.5–2.7 °C). This seemed reasonable since, under RCP8.5, GHG emissions continue to grow, in contrast to RCP4.5, in which emissions peak around 2040 and ultimately stabilize by 2080 [20,59]. Note that the min or max models predicted lower or higher temperature rise, respectively, compared to the median model since we developed those models based on the minimum, median and maximum projections of the monthly temperature changes (see Section 2.2.2 for details). The inter-model variations (difference between the minimum and maximum predicted temperature increase) range between 0.7–3.3 °C, which characterizes the uncertainties in climate modeling, as shown in Figure 3.

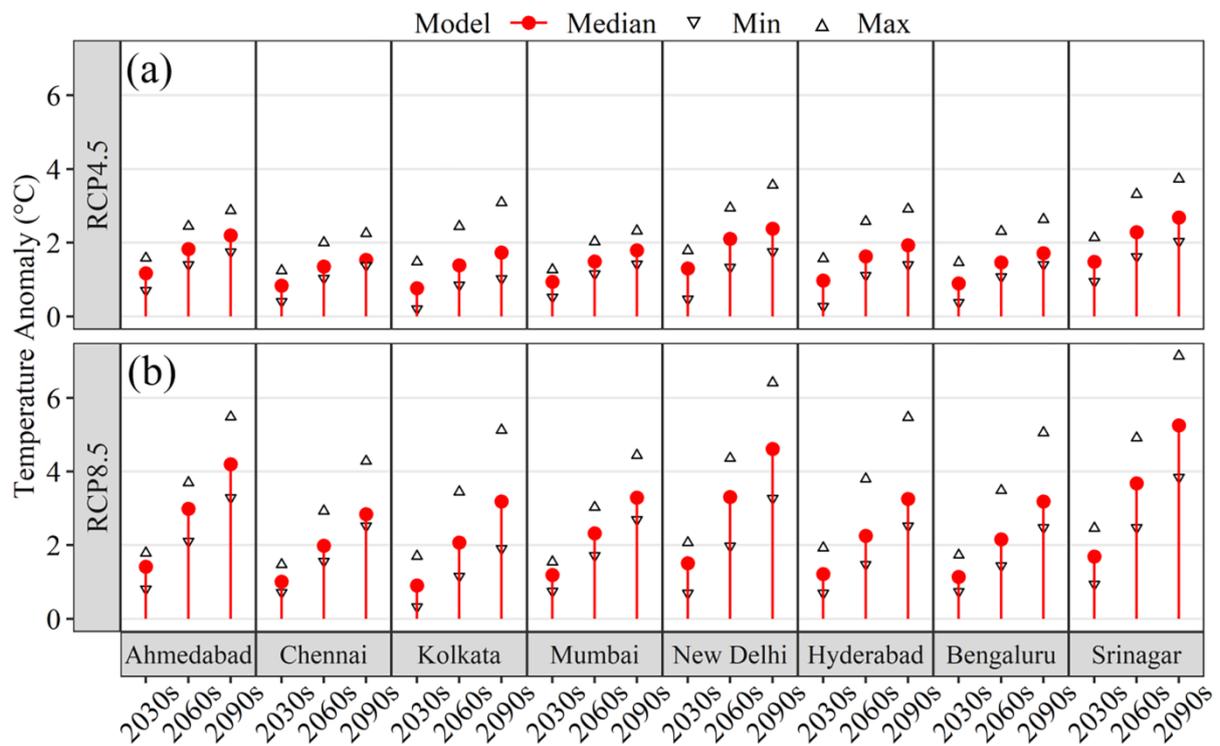

*Figure 3: Increase in the annual mean temperature by the 2030s, 2060s, and 2090s compared to the 1990s in major Indian cities under a) RCP4.5 and b) RCP8.5 scenarios.*

Due to rising temperatures, the annual-averaged values of specific humidity (mass of water vapor divided by the mass of dry air) are also projected to increase over their baseline values, as shown in Figure 4. The specific humidity will rise by 5–11 % in the 2030s, 8–24 % in the 2060s, and 10–36 % in the 2090s, according to the median model, depending on the city and the emission scenario. Note that the minimum and maximum changes in specific humidity values do not correspond to the min and max models since temperature was the parameter used in deriving these models, as described in Section 2.2.2.

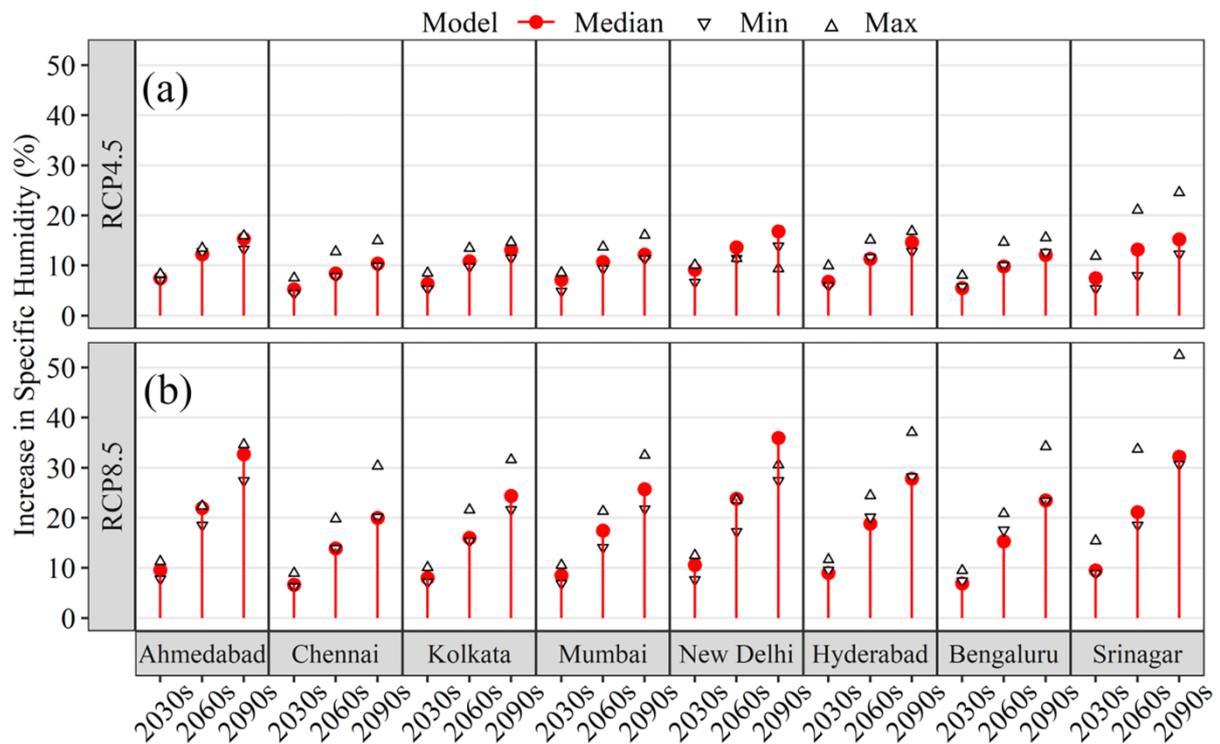

*Figure 4: Increase in the annual mean specific humidity by the 2030s, 2060s, and 2090s compared to the 1990s in major Indian cities under a) RCP4.5 and b) RCP8.5 scenarios.*

Our median model also showed that relative humidity would generally increase in the future for all cities except Srinagar (see Figure SM3). By the 2090s, RH will be 0.5–3.3 percentage points higher than its current value under both emission scenarios, consistent with IPCC predictions [60]. The RH trends obtained by the min model were qualitatively similar to the median model prediction; however, the max model predicted a decrease in future RH values (see Figure SM3). Thus, RH will likely increase in Indian cities, which sharply contrasts with decreasing RH trends (both current and future predictions) in most other parts of the world. Therefore, in the future, Indians will experience a considerable deterioration in outdoor thermal comfort due to rising temperature and RH levels.

The annual-averaged GHI is generally predicted to decrease slightly for every Indian city in the future under both emission scenarios, as shown in Figure SM4. The decrease in GHI, known as solar dimming, will be due to the increasing concentration of dust and aerosols in the atmosphere as a result of anthropogenic emissions [61]. Our results indicate that GHI would consistently decrease in future years due to a steady rise in anthropogenic emissions under RCP8.5. However, under RCP4.5, the GHI first falls (in the 2030s) and then rises slightly in the future periods, probably because anthropogenic emissions decline after 2040 in this scenario. Our results are consistent with a previous study on future GHI prediction in India [62]. Since the changes in GHI are relatively small (−3 % to 7 %), they would not significantly affect the buildings' H/C energy consumption. We also found that the projected annual-averaged wind speed did not show any trends, and its changes were within −6 % to 12 % (see Figure SM5), which would also not have much impact on the building energy consumption.

Overall, rising temperature and RH levels due to climate change will lead to a considerable deterioration in people's thermal comfort, necessitating the need for more cooling energy in India. The other climate variables, like GHI and wind speed, do not seem to change significantly to affect thermal comfort or cooling energy consumption.

### 3.2. Current and future H/C energy demand

We conducted 152 simulations under current and future climatic conditions to quantify the changes in H/C energy consumption for the selected cities, including 8 baseline simulations (for the 1990s) and 144 simulations for the FPs (8 cities × 3 FPs × 2 RCPs × 3 climate models). Furthermore, we conducted additional 456 simulations to quantify the contribution of each load component (walls, windows, infiltration, and internal) to the total H/C energy consumption, as explained in Section 2.1.3. Those results are presented in the following subsections.

### 3.2.1. Cooling energy demand

Table 3 shows the cooling system's annual usage hours and the corresponding energy consumption of the archetype room for the baseline period (the 1990s). The cities in climate zones: warm-humid, hot-dry, and composite, had significant cooling system usage and energy requirements. On the other hand, cities experiencing temperate (Bengaluru) or cold (Srinagar) climates had relatively low cooling energy needs. The latent cooling energy was between 5–24 % of the total, with cities having warm and humid climates (Chennai, Kolkata, and Mumbai) requiring maximum latent cooling.

*Table 3: Cooling system's annual usage hours and energy consumption for the archetype room in major Indian cities in the 1990s.*

| City | System usage in the 1990s (h) | Cooling energy (kWh) in the 1990s | | |
|---|---|---|---|---|
| | | Sensible (% of total) | Latent (% of total) | Total |
| Ahmedabad | 2931 | 1980 (86 %) | 336 (14 %) | 2315 |
| Bengaluru | 3111 | 1100 (89 %) | 133 (11 %) | 1233 |
| Chennai | 3650 | 2410 (78 %) | 659 (22 %) | 3068 |
| Hyderabad | 3308 | 1628 (89 %) | 191 (11 %) | 1818 |
| Kolkata | 2807 | 1653 (76 %) | 509 (24 %) | 2162 |
| Mumbai | 3567 | 1851 (79 %) | 492 (21 %) | 2343 |
| New Delhi | 2379 | 1678 (85 %) | 277 (14 %) | 1955 |
| Srinagar | 1318 | 491 (95 %) | 23 (5 %) | 514 |

Figure 5 shows the projected changes in the room's annual cooling energy consumption by the 2030s, 2060s, and 2090s over the 1990s for the different cities. By the 2030s, the increase in cooling energy will be about the same (187–487 kWh or 11–40 %, depending on the city) for both RCPs, as per the median model. However, by the 2090s, the increase in cooling demand would be substantially higher in RCP8.5 (797–1504 kWh or 42–155 %) than in RCP4.5 (377–760 kWh or 23–73 %). Note that the cooling energy requirements are projected to increase consistently from the 1990s to the 2090s under both emission scenarios for all cities, irrespective of the model (min, median, or max) used for making future projections. The inter-model variability (difference between the min and max model predictions of cooling energy) ranges between 166–1140 kWh (9–118 percentage points) as shown in Figure 5, demonstrating the significant uncertainties associated with the climate model predictions.

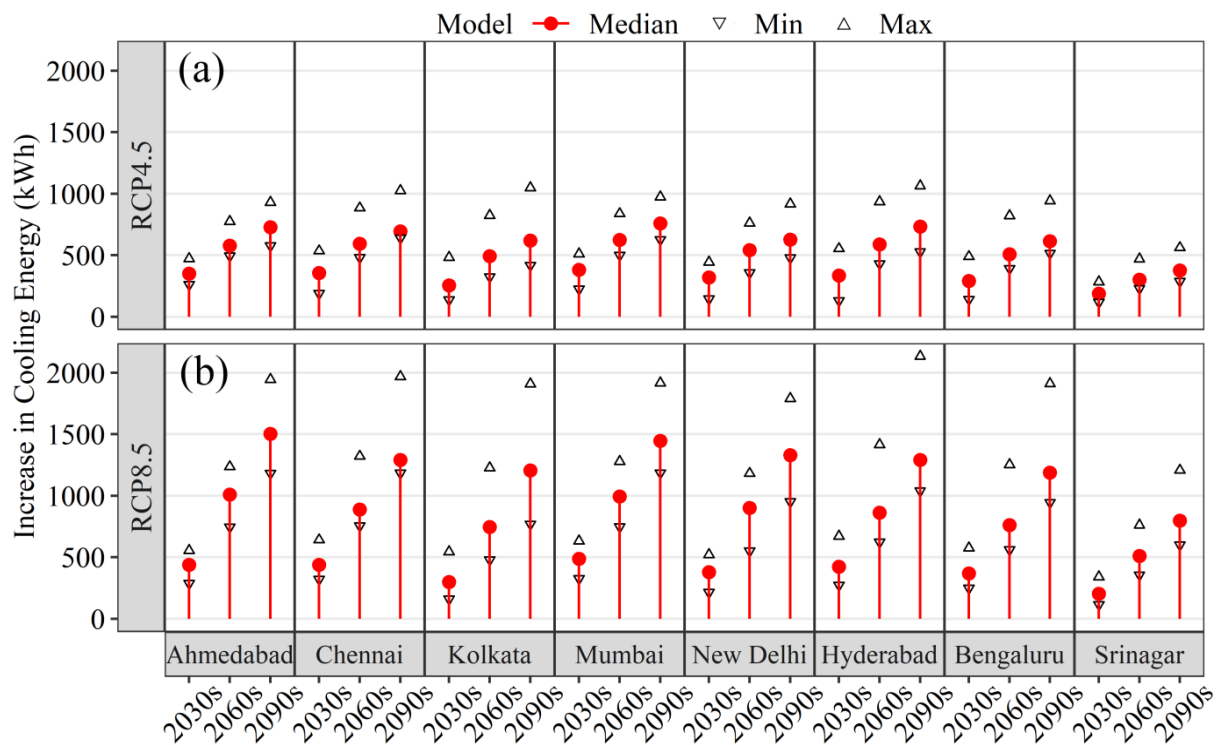

*Figure 5: Increase in the annual cooling energy consumption of the archetype room by the 2030s, 2060s, and 2090s compared to the 1990s in major Indian cities under a) RCP4.5 and b) RCP8.5 scenarios.*

### 3.2.2. Cooling load components

We also quantified the individual cooling load components (walls, windows, infiltration, and internal) for the archetype room using the methodology described in Section 2.1.3. The load components are shown in Figures 6 a–d for one representative city from those climate zones with significant cooling needs (other cities shown in Figures SM6 a–d) from the 1990s to the 2090s under both emission scenarios using the median model. We only discuss results obtained from the median model here since the results obtained from the min and max models (shown in Figures SM7 a–p) were generally similar.

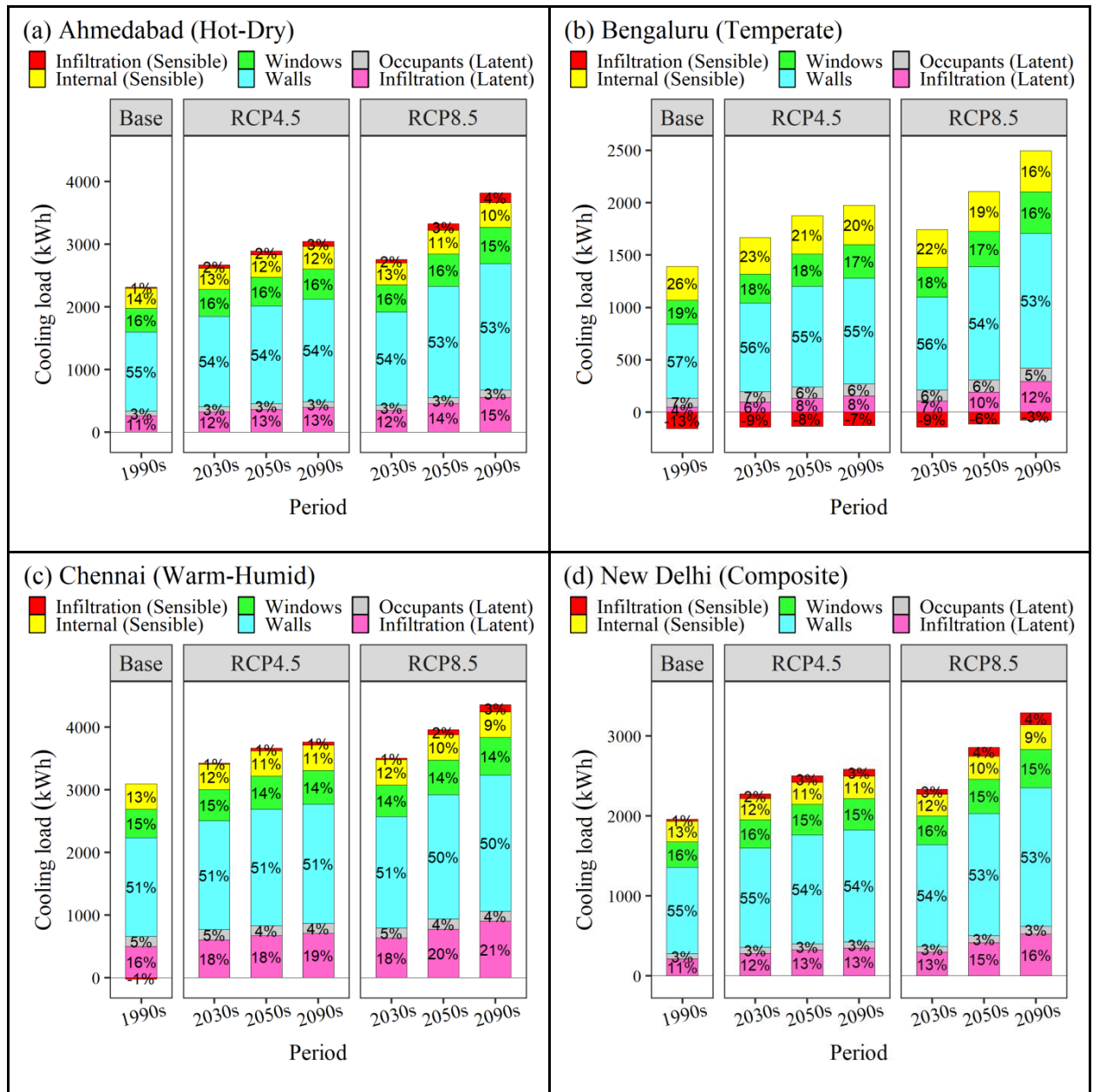

*Figure 6: Baseline and median model projections of the archetype room's cooling load components under RCP4.5 and RCP8.5 scenarios in a) Ahmedabad, b) Bengaluru, c) Chennai, and d) New Delhi.*

Figures 6 and SM6 show that the two *exposed walls* of the room contributed the maximum towards the cooling load, i.e., between 48–71 % in the 1990s, depending on the city. Although the wall cooling load would substantially increase in the future due to rising outdoor temperatures, its contribution would either remain the same or slightly decrease (by 1–7 percentage points). Figures 6 and SM6 also show that the two *windows* were generally the next highest contributor to the cooling load, which ranged between 15–19 % of the total load in the 1990s. Like the wall load, the window cooling load's proportional contribution would either remain constant or slightly decrease (by 1–3 percentage points) in future years for each city under both emission scenarios.

The proportion of *latent infiltration load* was between 1–18 % for all cities in the 1990s (see Figures 6 and SM6). The share of latent infiltration load was highest in the three warm and humid cities (16–18 %), and lowest (1 %) in Srinagar due to its cold climate in the 1990s. In contrast to the structural cooling loads, the proportional contribution of latent infiltration load would increase slightly (by 1–8 percentage points) in future years for each city under both emission scenarios. The contribution of *sensible infiltration load* was almost negligible in hot-dry, warm-humid, and composite climate zones in the 1990s (less than 1 %). Whereas infiltration currently (the 1990s) provides sensible cooling (13–22 %, shown as negative values in Figures 6 and SM6) in Bengaluru (temperate) and Srinagar (cold) due to low nighttime temperatures. The proportional contribution of sensible infiltration load would increase (by 1–14 percentage points) in future years for each city under both emission scenarios. There would also be a significant reduction in the nighttime cooling provided by infiltration in Bengaluru and Srinagar (by 4–14 percentage points). Overall, rising temperature and humidity levels significantly increase (by 2–19 percentage points) the contribution of infiltration loads (both sensible and latent) in the future and also reduce the potential of natural ventilation for free nighttime cooling.

The contribution of *sensible and latent internal loads* ranged between 13–27 % and 1–7 %, respectively, for all cities in 1990s, as shown in Figures 6 and SM6. The proportional contribution of sensible internal load would decrease (by 1–11 percentage points) in future years for each city under both emission scenarios since the sensible internal load almost remains constant while the other load components drastically increase. Similarly, the contribution of latent internal load would also slightly decrease (less than 3 percentage points) in future years for each city.

Overall, our results show that the structural load components (walls and windows) currently contribute and would continue to contribute the maximum to the cooling load (more than 60 % of the total) in all cities, and should be the primary target for energy-efficiency measures. Infiltration is the next significant contributor to the cooling load (2–17 %) for cities belonging to the hot-dry, warm-humid, and composite climate zones, whose contribution would further increase in the future. However, reducing air infiltration might not be a feasible strategy to reduce energy consumption in those cities since the assumed infiltration rate (0.75 ACH = 9.5 L/s) was already quite low, and a lower value may create indoor air quality concerns. For cities in temperate and cold climate zones, infiltration currently provides cooling (negative contribution to the load), but this effect would significantly decrease in the future, pointing towards the declining scope of natural ventilation.

### 3.2.3. Heating energy demand

This section discusses the heating demand in baseline and future periods for Srinagar (cold) and New Delhi (composite) because only those cities have significant heating requirements. In the 1990s, Srinagar had high heating system usage hours (1,262 hours) and energy requirements (511 kWh), while New Delhi had relatively low system usage hours (308 hours) and energy needs (63 kWh) since it has a warmer climate than Srinagar. Figure 7 shows the decrease in the archetype room's annual heating demand for Srinagar and New Delhi in the 2030s, 2060s, and 2090s over the 1990s under RCP4.5 and RCP8.5. In the 2030s, the decrease in heating energy demand will be almost the same (30–154 kWh or 24–49 %, depending on the city) for both RCPs, as predicted by the median model. However, by the 2060s and 2090s, heating requirements decrease much more under RCP8.5 (58–369 kWh or 58–100 %) than under RCP4.5 (45–232 kWh or 38–76 %). The inter-model variability (difference between the min and max model predictions of heating energy) ranges between 8–147 kWh (13–57 percentage points), as shown in Figure 7.

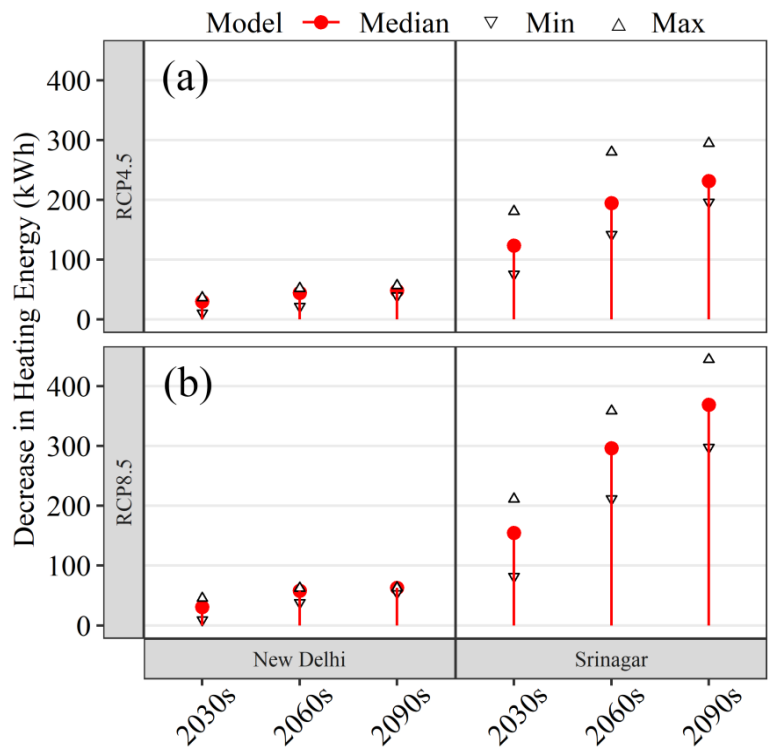

*Figure 7: Decrease in the annual heating energy consumption of the archetype room by the 2030s, 2060s, 2090s compared to the 1990s in New Delhi and Srinagar under a) RCP4.5 and b) RCP8.5 scenarios.*

### 3.2.4. Heating load components

Subsequently, we quantified the heating load components for Srinagar and New Delhi by performing a similar analysis as carried out for the cooling load components. Once more, we report the results obtained with the median model only since the min and the max models also generated qualitatively similar results (see Figures SM8 a–d).

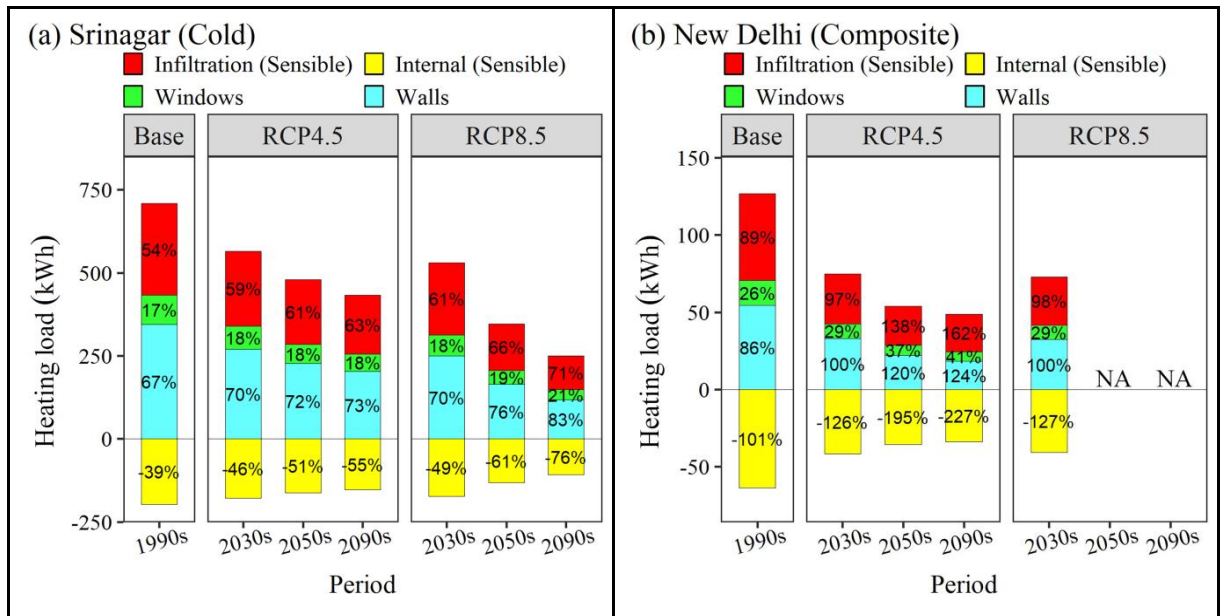

*Figure 8: Baseline and median model projections of the archetype room's heating load component under RCP4.5 and RCP8.5 scenarios in a) Srinagar and b) New Delhi.*

Figures 8 a–b show that in both cities, the exposed walls and infiltration had the most contribution to the heating energy requirement in the 1990s, whereas windows had the least. On the other hand, internal heat gains (occupants and lights) provided significant heating to the room and reduced the heating requirements (shown with negative values in Figure 8). The load components were not calculated for New Delhi in the 2060s and 2090s under the RCP8.5 scenario since the heating load was negligible under those conditions. The load components would continue to have qualitatively similar contributions to the total heating load in both cities in future years under both emission scenarios, with exposed walls and infiltration accounting for the largest share. Thus, insulating external walls seems to be an effective way to reduce current and future heating energy demand in Indian cities. The other alternative, reducing infiltration, may not be viable since it will lead to air quality concerns, as discussed in Section 3.2.2.

### 3.3. Study limitations and recommendations for future research

The study provides a detailed assessment of the impact of climate change on the heating and cooling load components of an archetypical residential room in major Indian cities. We conducted energy simulations for an archetypical room, instead of modeling prototype buildings (buildings possessing similar characteristics) that can represent the complete residential stock [63,64] since such prototype models have not been developed for India. Future research should focus on studying the composition of India's residential building stock, occupancy conditions, and H/C cooling energy usage patterns to develop prototype buildings that can aid building retrofitment programs and energy-efficiency solutions. In the absence of building prototypes, we modeled a bedroom with its own H/C system in a multi-family apartment since centralized H/C is almost non-existent in Indian residences, and even affluent Indian households typically use room H/C systems only [43]. Thus, our study provides important insights into reducing the H/C energy consumption of wealthy urban households but does not consider the diversity of India's residential building stock, which should be studied in future investigations. This research also did not consider the effect of building orientation, occupancy patterns, operating conditions of the H/C system, and energy-efficiency measures, which can be explored in future research.

**4. Conclusions**

This investigation conducted a multi-model and multi-scenario assessment of the impact of climate change on the H/C load components (walls, windows, infiltration, occupants, and lighting) of an archetypical residential room in eight Indian cities. Using a novel approach, we estimated the load components for an archetype bedroom (3.33 m × 4.03 m × 3.18 m in size), as it is prevalent in India to use H/C in bedrooms only, and identified those building elements that should be targeted for improving energy efficiency. The study led to the following conclusions:

- By the end of this century, the annual average temperature in Indian cities will rise by 1.5–5.3 °C, and the relative humidity will increase by 0.5–3.3 percentage points, according to the median climate model predictions. Increasing temperature and relative humidity conditions will significantly deteriorate people's thermal comfort, necessitating the need for more cooling.
- The cooling energy demand of the archetype room will increase by 187–487 kWh (11–40 %) by the 2030s and by 377–1504 kWh (23–155 %) by the 2090s compared to the 1990s, depending on the location and emission pathway, according to the median model. The cooling energy requirements are projected to increase consistently from the 1990s to 2090s under both emission pathways for all cities, irrespective of the climate model (min, median, or max) used for future projections. However, considerable variations were found in the magnitude of the increase, depending on the climate model and emission scenario considered.
- The structural load components (walls and windows) currently contribute and would continue to contribute maximum to the cooling load (more than 60 % of the total) in all cities, and should be the primary target for energy-efficiency measures.
- In contrast to rising cooling needs, the room's heating energy demand will decrease by 30–154 kWh (24–49 %) by the 2030s and by 48–369 kWh (45–100 %) by the 2090s, depending on the city and emission pathway, as per the median model. Once again, significant inter-model and inter-pathway variabilities exist in future predictions, although the decreasing trend was consistent.

- External walls contributed the maximum to the heating needs (more than 67 %), and their share is expected to remain substantial in the future; thus, insulating them could effectively reduce the heating energy.

**Funding**

This work was partially supported by the Birla Institute of Technology and Science, Pilani, through its Outstanding Potential for Excellence in Research and Academics (OPERA) award. RSS and ACR also acknowledge the financial support extended by the Chandrakanta Kesavan Centre for Energy Policy and Climate Solutions at the Indian Institute of Technology Kanpur.